\documentclass{iaus}
\usepackage{graphicx}

\title[Gallery of PN Spectra] 
{Gallery of Planetary Nebula Spectra}

\author[Kwitter \& Henry]   
{Karen B. Kwitter$^1$
  \break \and Richard B.C. Henry$^2$}

\affiliation{$^1$Department ofAstronomy, Williams College, Williamstown, MA 01267 USA
 \break email: kkwitter@williams.edu\\[\affilskip]
$^2$H. L. Dodge Department of Physics \& Astronomy, University of
Oklahoma, \break Norman, OK 73019 \break email: henry@nhn.ou.edu}

\pubyear{2006}
\volume{234}  
\pagerange{1-2}
\date{?? and in revised form ??}
\jname{Planetary Nebulae in our Galaxy and Beyond}
\editors{M.J. Barlow \& R.H. M\'endez, eds..}
\begin{document}

\maketitle

\firstsection 
\section{Overview}

	In the course of our abundance studies over the past decade we have accumulated more than 120 high-quality, medium resolution spectra of planetary nebulae (PNe) from 3600-9600$\AA$  using the KPNO 2.1m Goldcam CCD spectrograph and the CTIO 1.5m RC spectrograph. Results have been published in, {\it e.g.}, \cite[Kwitter \& Henry (1998)]{Kwitter98}; \cite[Henry, Kwitter \& Balick (2004)]{Henry04}; and \cite[Milingo et al. (2006)]{Milingo06}. We have created this website as a place where the spectra are available for graphical display, and where PN atlas information and image links are tabulated. The displayed spectra are {\bf not} research quality;  they are samples for display and educational purposes only ({\it e.g.}, strong lines may be saturated). Investigators interested in the original data should contact the authors.

\section{The Website:  {\it \bf http://oit.williams.edu/nebulae}} \label{sec:website}

\subsection{Features}
The website's contents are displayed in a sidebar. The {\it Home} page describes the site and lists our publications relating to this data. The {\it Introduction} page is a short course on PNe; the {\it Help} and {\it Contact} pages are self-explanatory. The {\it Browse} page  is the heart of the website and contains the master PN table, listing all 128 objects, relevant atlas data, and links to images. The master PN table can be re-ordered by object name, Peimbert type, galactic or equatorial coordinates, distance from the sun, the galactic center, or the galactic plane. For each object, a window can be opened wherein the object's spectrum is displayed in a zoomable sub-window. Line identification templates are provided. A sidebar contains links to navigate within the site. We have written two exercises  to introduce  beginning astronomy students to the wealth of  information that PN spectra contain. These exercises are completely self-contained and are designed for students from high school to college.

\subsection{Exercises}\label{sec:exercises}

{\it Exercise 1: Emission Lines and Central Star Temperature in Planetary Nebulae:} Students examine a set of three PNe to search for emission lines from various highly-ionized atoms. By considering the ionization potential of each of these ions, the students can then place the three PNe in order of the temperature of the central star. 

{\it Exercise 2: Detecting Interstellar Reddening:} Students examine a set of eight PNe and measure the relative heights of H$\alpha$ and H$\beta$ as indicators of their relative observed fluxes. Students are then asked to describe the trend of the observed ratio with galactic latitude, and to infer that the distribution of dust in the Galaxy is concentrated in the disk.

We plan to add two more exercises: one on determining nebular density using the  [S II] $\lambda\lambda$6717/6731 ratio, and another examining  the Paschen lines. We will also be looking into alternative Java applets for displaying and measuring the spectra.

\begin{acknowledgments}
We are grateful to NSF for support under grants AST-9819123 and AST-0307118, and to Williams College and U. Oklahoma. We also thank the Williams Instructional Technology summer intern program and many individual Williams and Oklahoma undergraduates. KBK and RBCH (plus J.B. Milingo) were Visiting Astronomers at Kitt Peak and Cerro Tololo, NOAO, which are operated by the Association of Universities for Research in Astronomy, Inc. (AURA) under cooperative agreement with the National Science Foundation.
\end{acknowledgments}

\end{document}